\documentclass[pra,twocolumn,aps,superscriptaddress,showpacs,floatfix]{revtex4}
\usepackage{amssymb}
\usepackage{graphicx}
\usepackage{dcolumn}
\usepackage{bm}
\usepackage{amsmath}
\usepackage{amsbsy}
\usepackage{amssymb}
\usepackage{graphicx}
\usepackage{color}
\usepackage{epstopdf}
\usepackage{mathrsfs}
\usepackage{multirow}
\usepackage[colorlinks,linkcolor=magenta,citecolor=blue,urlcolor=blue]{hyperref}
\usepackage{etoolbox}

\begin{document}

\title{Floquet analysis of extended Rabi models based on high-frequency expansion}
\author{Yibo Liu}
\affiliation{
{Institute of Theoretical Physics and State Key Laboratory of Quantum Optics and Quantum Optics Devices, Shanxi University, Taiyuan 030006, China}
}

\author{Lijun Mao}
\affiliation{
{Department of Physics, Taiyuan Normal University, Jinzhong 030619, China, Taiyuan 030006, China}
}

\author{Yunbo Zhang}
\email{ybzhang@zstu.edu.cn}
\affiliation{
{Department of Physics and Key Laboratory of Optical Field Manipulation of Zhejiang Province, Zhejiang Sci-Tech University, Hangzhou 310018, China}
}

\begin{abstract}
The extended quantum Rabi models make a significant contribution to understand the quantum nature of the atom-light interaction. We transform two kinds of extended quantum Rabi model, anisotropic Rabi model and asymmetric Rabi model, into rotating frame, and regard them as periodically driven quantum systems. The analytical solutions of the quasi-energy spectrum as well as the Floquet modes for both models are constructed by applying the Floquet theory and the high-frequency expansion, which is applied to the non-stroboscopic dynamics of physical observables such as atomic inversion, transverse magnetization, atom-field correlation, etc. For anisotropic Rabi model, the quasi energy fits well with the numerical results even when the rotating-wave coupling is in the deep-strong coupling regime $g \simeq \hbar \omega$ if the counterrotating terms is small enough compared to the driving frequency. Avoided level crossing may occur for quasi energy with the same parity when the positive branch spectrum lines for the total excitation number $N$ cross the negative branch lines for $N+2$, while the high frequency expansion fails to predict this due to the conservation of the total excitation number. Furthermore, we present analytical and numerical study of the long-time evolution of population and figure out the analytical method is credible for the population dynamics. For asymmetric Rabi model, we find that the external bias field which breaks the parity symmetry of total excitation number tends to cluster the upper and lower branches into two bundles, and the detuning induced gap in the first temporal Brillouin zone shows a quadratic dependence on the bias. The Fourier analysis is applied to extract the frequency composition and the two-frequency driving behavior is revealed. Varying the bias strength will change the time-averaged value of the oscillation, which shows how the bias competes with the detuning and atom-field coupling in the driving dynamics. Both models prove that treating the Hamiltonian in the rotating frame by Floquet theory gives an alternative tool in the study of interaction between atom and light.
\end{abstract}

\maketitle
\section{Introduction}
The Rabi model \cite{Rabi1937}, which describes a two-level atom interacting with a single-mode classical filed, plays an important role in understanding atom-field interaction. A solvable fully quantum-mechanics model, Jaynes-Cummings (JC) model \cite{Jaynes1963}, gives the general and basic physics of quantum Rabi model in the rotating-wave approximation (RWA) \cite{Gerry2005}. The development of experiments in circuit quantum electrodynamics \cite{Niemczyk2010,Crespi2012,Gely2017,Mezzacapo2014,Yoshihara2016}, 2D electron gases \cite{Hagenmuller2011,Smolka2014}, and trapped ions \cite{Pedernales2015,ChengXH2018,LvD2018,CaiML2021}, etc, has already driven the light-matter interaction into the strong coupling regime where RWA is not applicable and the counter-rotating wave terms (CRTs) can not be neglected. More and more generalized quantum Rabi model were proposed to study different kinds of interaction beyond RWA, e.g. Rabi-Stark model \cite{Eckle2017,XieYF2019a,XieYF2019b,ChenXY2020}, Dicke model \cite{Dicke1954,Garraway2011,Kirton2019}, Buck-Sukumar model \cite{Cordeiro2007,Rodriguez-Lara2014}, anisotropic Rabi model \cite{ZhangG2015,XieQT2014,XieQ2017,WangGC2019,Skogvoll2021}, and asymmetric Rabi model \cite{XieQ2017,Ashhab2020,ChenQH2012,LiZM2021,Reyes-Bustos2021,Braak2011,Yoshihara2017}, etc. It's thus necessary to find a proper treatment of CRTs to make the extended quantum Rabi models solvable. Numerical and analytical approaches were presented to obtain the energy spectrum of these models, such as $G-$function \cite{Braak2011}, Bogoliubov operators \cite{ChenQH2012}, and generalized rotating wave approximation \cite{Irish2007,ZhangYY2016}, etc. Very recently a quantum Rabi triangle system has been proposed as an elementary building block to explore the nature of emerging quantum many-body phases \cite{ZhangYY2021}. An adiabatic scheme for the fast and deterministic generation of a two-qubit Bell state and arbitrary single-photon multimode W states was proposed based on one-photon solutions to the multiqubit multimode quantum Rabi Model \cite{PengJ2021}. The quantum phase transition has been observed in the experiments with a single $^{171}{\rm Yb}{^+}$ ion confined in a linear Paul trap \cite{LvD2018,CaiML2021}. An experimental scheme with a transmon qubit capacitively coupled to a LC resonator has been proposed to implement the anisotropic quantum Rabi model in a circuit quantum electrodynamics system via periodic frequency modulation \cite{WangGC2019}. A magnon-spin-qubit ensemble in which a spin qubit exchange-coupled to an anisotropic ferromagnet is suggested to physically realize the quantum Rabi model from the isotropic to the Jaynes-Cummings limit with coupling strengths that can reach the deep-strong regime \cite{Skogvoll2021}. By inductively coupling a flux qubit and an LC oscillator via Josephson junctions, superconducting qubit-oscillator circuits in the deep strong coupling regime has been realized with a flux bias \cite{Yoshihara2017}.

The Floquet theory is generally applied in the periodic driven quantum system to gain the nontrivial physical properties \cite{Rahav2003,Goldman2014,Eckardt2015,Bukov2015,Bukov2014,Kohler2017,Oka2019,Rechtsman2013,Else2016,Lindner2011,Shirley1965,Casas2001}. According to the theory, the time-evolution operator from initial time $t_i$ to the final time $t_f$ can be written as
\begin{align}
	\hat{U}(t_f,t_i) = {\rm e}^{-{\rm i}\hat{K}(t_f)}{\rm e}^{-\frac{\rm i}{\hbar}\hat{H}_{F}\cdot (t_f-t_i)}{\rm e}^{{\rm i}\hat{K}(t_i)},
\end{align}
where $\hat{H}_{F}$ is the time-independent Floquet Hamiltonian to describe the long-time evolution and $\hat{K}(t)$ is the kick operator to describe the short-time behavior. There are two choices of the description, the stroboscopic and the non-stroboscopic dynamics. The former concludes that both the stroboscopic Floquet Hamiltonian $\hat{H}_{F}[t_0]$ and the stroboscopic kick operator $\hat{K}_F[t_0](t)$ depend on the Floquet gauge $t_0$, which is defined as the time where the first period begins. A very efficient tool to compute the stroboscopic Floquet Hamiltonian in the high-frequency limit is the the Magnus expansion, which is a perturbative scheme in the inverse driving frequency $1/\Omega$ to compute $\hat{H}_{F}[t_0]$ \cite{Bukov2015,Blanes2009}. It is specifically suitable to use this stroboscopic dynamics to describe the system in the simplest case when the initial time and final time are fixed on $t_0$ and $t_0+nT$ with the driving period $T=2\pi /\Omega$, as the stroboscopic kick operator in this case reduces to the zero. This method is widely applied in Floquet engineering of many-body localization \cite{Abanin2016}, counterdiabatic protocols \cite{Kuwahara2016}, and generic transient dynamics \cite{Claeys2019} in quantum many-body system. The other one is the non-stroboscopic dynamics which is described by the $t_0$-independent effective Floquet Hamiltonian $\hat{H}_{\rm eff}$ and the non-stroboscopic kick operator $\hat{K}_{\rm eff}(t)$. This approach offers the advantage that the dependence on the Floquet gauge will not enter the inverse-frequency expansion of $\hat{H}_{\rm eff}$ \cite{Goldman2014,Bukov2015}. If one is interested in Floquet non-stroboscopic dynamics, in current and linear response, or the spectral properties of the Floquet Hamiltonian, then the effective description offers an advantage, since it gives a Hamiltonian which does not contain terms that depend on the phase of the drive. Non-stroboscopic dynamics is capable of capturing the evolution governed by the Floquet Hamiltonian of any observable associated with the effective high-frequency model \cite{Bukov2014,Kohler2017,Oka2019}. 

In recent years, some Rabi-type models has been studied by applying Floquet theory \cite{LeeTE2015,XieQ2018,Dasgupta2015,Bastidas2012,WangYF2021,DuanLW2020}, including engineering the non-Hermitian Hamiltonian with semiclassical Rabi models and driving fully quantum Rabi-type models with time periodical parameters. Light in its nature is temporal periodic and the Rabi model is a periodically driven system in the first place. But the application of the Floquet theory directly on the quantum Rabi models is rare, as quantum optics description utilizing the field quantization maps the atom-field interaction into a time independent Hamiltonian - the field creation and annihilation operators does not depend on time in Schr\"odinger picture. The analysis of the quasienergy and the dynamic evolution of Rabi model is expected to lead to different view of understanding the various kinds of atom-light interaction. Although the Floquet stroboscopic dynamics is sufficient for the time evolution, the non-stroboscopic effective model in many cases provides the analytical form of the quasi-energy spectrum and Floquet modes, which is applicable in the evaluation of dynamics of physical observables. 

In this paper, we consider two extended Rabi models, i.e. the anisotropic and asymmetric quantum Rabi models, and study their non-stroboscopic dynamics in the framework of Floquet theory. We deliberately break the symmetry in the standard Rabi model, in one case, the $U(1)$ symmetry in atom-field coupling - the rotating and counter-rotating interactions are governed by two different coupling constants, in the other case, the $Z_2$ parity symmetry of total excitation number - a bias field is applied on the transverse direction. In the rotating frame the Hamiltonian can be regarded as a periodic driving in the interaction picture. In Section II, we apply the Floquet theory and high-frequency expansion to the anisotropic Rabi model. The quasi-energy spectrum and population dynamics derived from the time-independent effective model will be investigated and compared with the numerical result in the extended Floquet Hilbert space. In Section III, we carry out the similar procedure on the asymmetric Rabi model, and the driving dynamics of some physical observables will be analyzed. By comparing the numerical result and the effective model, we aim to find the parameter regime for the application of the high-frequency expansion. We conclude our results in Section IV and the details of the the expansion of the effective Hamiltonian and the numerical scheme for quasi-energy spectrum are presented in the appendix.

\section{Anisotropic Rabi model}
\subsection{Formalism and high-frequency expansion}
Our first model is the anisotropic Rabi Model(AiRM) \cite{XieQT2014,XieQ2017}, which can be described by the Hamiltonian in lab frame
\begin{eqnarray}
\hat{H}_{\rm{AiRM}} &=& \frac{1}{2}\hbar\omega_{0}\hat{\sigma}_{z}+\hbar\omega\hat{a}^{\dag}\hat{a}+g(\hat{a}^{\dag}\hat{\sigma}_{-}+\hat{a}\hat{\sigma}_{+}) \nonumber\\
&&+ g^{\prime}(\hat{a}^{\dag}\hat{\sigma}_{+}+\hat{a}\hat{\sigma}_{-}).
\label{oH1}
\end{eqnarray}
Here, $\hat{a}^{\dag}$ and $\hat{a}$ are the creation and annihilation operators for photons of single-mode frequency $\omega$, $\hat{\sigma}_+ =(\hat{\sigma}_x+{\rm i}\hat{\sigma}_y)/2$ and $\hat{\sigma}_-= (\hat{\sigma}_x-{\rm i}\hat{\sigma}_y)/2$ are the atomic transition operators, and $\hat{\sigma}_i (i=x,y,z)$ are the Pauli matrices of the atom  with the level difference characterized by the frequency $\omega_0$. $g$ is the coupling strength between the atom and the field of the RWTs, while $g^{\prime}$ is the coupling strength of CRTs. Clearly, when $g^\prime = 0$, the AiRM reduces to the JC model with rotating wave approximation, while for $g^{\prime} = g$, it becomes the standard quantum Rabi model. So this is an appropriate candidate to study the effect caused by CRTs.

To apply the Floquet theory, one needs the Hamiltonian to be time-dependent and temporal periodic. There is a convenient way to achieve this, i.e. putting the Hamiltonian in the rotating frame. After the time-dependent gauge transformation by the unitary operator $\hat{V}(t) = \exp[-{\rm i}\omega(\hat{a}^{\dag}\hat{a}+\hat{\sigma}_z/2)t]$, the Hamiltonian in the rotating frame has the form
\begin{align}
\hat{H}_{\rm{AiRM}}^{\rm rot} (t)&= \hat{V}(t)^{\dag}\hat{H}_{\rm{AiRM}}\hat{V}(t)+{\rm i}\hbar{\frac{\partial\hat{V}^{\dag}(t)}{\partial t}}\hat{V}(t)\notag\\
&=\frac{1}{2}\hbar\Delta\hat{\sigma}_{z}+g(\hat{a}^{\dag}\hat{\sigma}_{-}+\hat{a}\hat{\sigma}_{+})\notag\\
&+g^{\prime}({\rm e}^{{\rm i}2\omega t}\hat{a}^{\dag}\hat{\sigma}_{+}+{\rm e}^{-{\rm i}2\omega t}\hat{a}\hat{\sigma}_{-}).\label{Hrot}
\end{align}
Comparing it with the Hamiltonian (\ref{oH1}) in the original gauge, we find the free field term has been eliminated and the field information is entangled with the atom. The RWTs remain time-independent in the new gauge, while the CRTs have the time dependence with the frequency $\Omega = 2\omega$, which can be regarded as a periodic driving. The time-independent part consists the JC model without the free field Hamiltonian, and the atomic level difference is characterized by the detuning between the photon and atom $\Delta = \omega_0 - \omega$. For high frequency driven system one needs that the photon frequency $\omega$ is large enough such that the atoms are in near resonance with the optical mode. 

The Floquet theory gives the guidance of solving these periodic driven cases in the high-frequency limit, i.e. the effective Hamiltonian can be expanded in order of $\Omega^{-n}$ as $\hat{H}_{\rm eff} = \sum_{n=0}^{\infty}H_{\rm eff}^{(n)}$. The corresponding kick operator can also be expanded as $\hat{K}_{\rm eff}(t) = \sum_{n=1}^{\infty}K_{\rm eff}^{(n)}(t)$. The first few terms of high-frequency expansion are calculated as (See Appendix A for details)
\begin{align}
\hat{H}^{(0)}_{\rm eff} &= \frac{1}{2}\hbar\Delta\hat{\sigma}_{z}+g(\hat{a}^{\dag}\hat{\sigma}_{-}+\hat{a}\hat{\sigma}_{+}),\label{H0}\\
\hat{H}^{(1)}_{\rm eff} &= \frac{g^{\prime 2}}{\hbar\Omega}(\hat{a}^{\dag}\hat{a}\hat{\sigma}_z-\hat{\sigma}_-\hat{\sigma}_+),\label{H1}\\
\hat{H}^{(2)}_{\rm eff} &= -\frac{g^{\prime 2}}{(\hbar\Omega)^2}[\hbar\Delta (\hat{a}^{\dag}\hat{a}\hat{\sigma}_z-\hat{\sigma}_-\hat{\sigma}_+)+g(\hat{a}^{\dag}\hat{a}\hat{a}^{\dag}\hat{\sigma}_-+h.c.)].\label{H2}
\end{align}
The zeroth order term of expansion \eqref{H0}, which is exactly the time-independent part of Hamiltonian in the rotating frame \eqref{Hrot}, plays a major role in the effective Hamiltonian as we expected. The CRTs enter in the first order correction \eqref{H1} showing the dependence of the effective Hamiltonian on the coupling strength $g'$. The second order correction \eqref{H2} consists of two parts: apart from a term similar to the first order correction (dressed by the detuning $\Delta$), there appears a two-photon interaction process which depends on both coupling strengths of RWTs and CRTs. With the increase of the correction order, more multiple-photon process will be brought into the effective Hamiltonian. On the other hand, the first few terms of the corresponding effective kick operator can be evaluated as
\begin{align}
\hat{K}_{\rm eff}^{(1)}(t) &= \frac{g^{\prime}}{{\rm i}\hbar\Omega}(\hat{a}^{\dag}\hat{\sigma}_{+}{\rm e}^{{\rm i}\Omega t}-h.c.),\\
\hat{K}_{\rm eff}^{(2)}(t) &=  \frac{g^{\prime}}{{\rm i}(\hbar\Omega)^2} [(g\hat{a}^{\dag 2}\hat{\sigma}_z-\hbar\Delta\hat{a}^\dag\hat{\sigma}_+){\rm e}^{{\rm i}\Omega t}-h.c.].
\end{align}
We see that the zeroth-order term in the Floquet Hamiltonian \eqref{H0} is simply the time-averaged Hamiltonian, whereas the zeroth-order non-stroboscopic kick operator is identically zero.

\subsection{Quasi-energy and Floquet modes}
Notice that the effective Hamiltonian conserves the total number of excitation $\hat{N} = \hat{a}^{\dag}\hat{a}+(\hat{\sigma}_z+1)/2$ 
, which enlarges the $Z_2$ symmetry of the original Hamiltonian \eqref{oH1} to $U(1)$ symmetry. Therefore, it only couples pairs of states such as $|1\rangle = |n,+\rangle$ and $\ |2\rangle = |n+1,-\rangle$ where $n$ is the photon number and $|\pm\rangle$ denotes the excited and ground states of the atom. The matrix of the effective Hamiltonian in these basis is given by
\begin{widetext}
	\begin{equation}
		\hat{H}_{\rm eff} = \begin{bmatrix}
			\frac{\hbar\Delta}{2}+\frac{g^{\prime 2}}{\hbar\Omega}n-\frac{\hbar\Delta g^{\prime 2}}{(\hbar\Omega)^2}n&g\sqrt{n+1}-\frac{gg^{\prime 2}}{(\hbar\Omega)^2}(n+1)\sqrt{n+1}\\
			g\sqrt{n+1}-\frac{gg^{\prime 2}}{(\hbar\Omega)^2}(n+1)\sqrt{n+1}&-\frac{\hbar\Delta}{2}-\frac{g^{\prime 2}}{\hbar\Omega}(n+2)+\frac{\hbar\Delta g^{\prime 2}}{(\hbar\Omega)^2}(n+2)
		\end{bmatrix}.\label{matrix}
	\end{equation}
\end{widetext}

We should firstly consider a special case that the total excitation number is zero. There is only one state which forms an independent space from others with the quasi-energy and the eigenvector
\begin{align}
E_0 &= -\frac{\hbar\Delta}{2}-\frac{g^{\prime2}}{\hbar\Omega}+\frac{\hbar\Delta g^{\prime 2}}{(\hbar\Omega)^2},\\
|\Phi_0\rangle &= |0,-\rangle. \label{phi0}
\end{align}
It's straightforward to get the quasi-energy spectrum by solving the eigenvalue problem of the matrix \eqref{matrix} for the case of non-zero total excitation number. Expansion of $E_{n\pm}$ with respect to $(\hbar\Omega)^{-1}$ up to the second order can be written as
\begin{align}
	E_{n\pm}^{(0)} &= \pm\frac{\Omega_{R}}{2},\\
	E_{n\pm}^{(1)} &= \frac{g^{\prime 2}}{\hbar\Omega}\left[-1\pm\frac{\hbar\Delta(n+1)}{\Omega_{R}}\right],\\
	E_{n\pm}^{(2)} &=\frac{g^{\prime 2}}{(\hbar\Omega)^2}\left\{\hbar\Delta\mp\frac{[\hbar\Delta g^{\prime}(n+1)]^2}{\Omega_R^{ 3}}\right.\notag\\
	&\left.\pm\frac{(g^{\prime 2}-2g^2)(n+1)^2-(\hbar\Delta)^2(n+1)}{\Omega_{R}}\right\},
\end{align}
where $\Omega_R = \sqrt{(\hbar\Delta)^2+4g^2(n+1)}$ is so-called Rabi frequency of the quantum Rabi model. We can also obtain the normalized eigenvectors of the effective Hamiltonian in the combination of the basis chosen before,
\begin{align}
|\Phi_{n\pm}\rangle =\frac{1}{\sqrt{1+ C^2_{n\pm}}} (C_{n\pm}|n,+\rangle+|n+1,-\rangle),\label{phinpm}
\end{align}
where the coefficient up to the second order reads
\begin{align}
	C_{n\pm}^{(0)} &= \frac{\hbar\Delta\pm\Omega_R}{2g\sqrt{n+1}},\\
	C_{n\pm}^{(1)} &= \frac{1}{\hbar\Omega}\frac{g^{\prime 2}\sqrt{n+1}}{g}\left(1\pm\frac{\hbar\Delta}{\Omega_R}\right),\\
	C_{n\pm}^{(2)} &= \frac{1}{(\hbar\Omega)^2}\frac{g^{\prime 2}\sqrt{n+1}}{2g}\left\{-\hbar\Delta \mp\frac{(\hbar\Delta)^2}{\Omega_R}\right.\notag\\
	&\left.\pm\frac{8g^{\prime 2}g^2(n+1)^2}{\Omega_R^{3}}\right\},
\end{align}
and the Floquet modes in the rotating frame which form an orthonormal basis at time $t$ can be written as
\begin{align}
|\phi_0\rangle &= {\rm e}^{-{\rm i}\hat{K}_{\rm eff}(t)}|\Phi_0\rangle,\\
|\phi_{n\pm}\rangle &= {\rm e}^{-{\rm i}\hat{K}_{\rm eff}(t)}|\Phi_{n\pm}\rangle.
\end{align}
We can easily check that in the case $g^{\prime} = 0$ the effective model reduces to the JC model. The quasi-energy is left with the zeroth order term and the system is characterized by the energy scale $\Omega_R$ and the driving frequency $\Omega$. This gives exactly the JC energy eigenvalues when rotating back to the lab frame. The kick operator $\hat{K}_{\rm eff}(t)$ becomes zero and the Floquet modes become time-independent and reduce to the eigenstates of the JC Hamiltonian.

\begin{figure}[tbp]
	\includegraphics[width=0.5\textwidth]{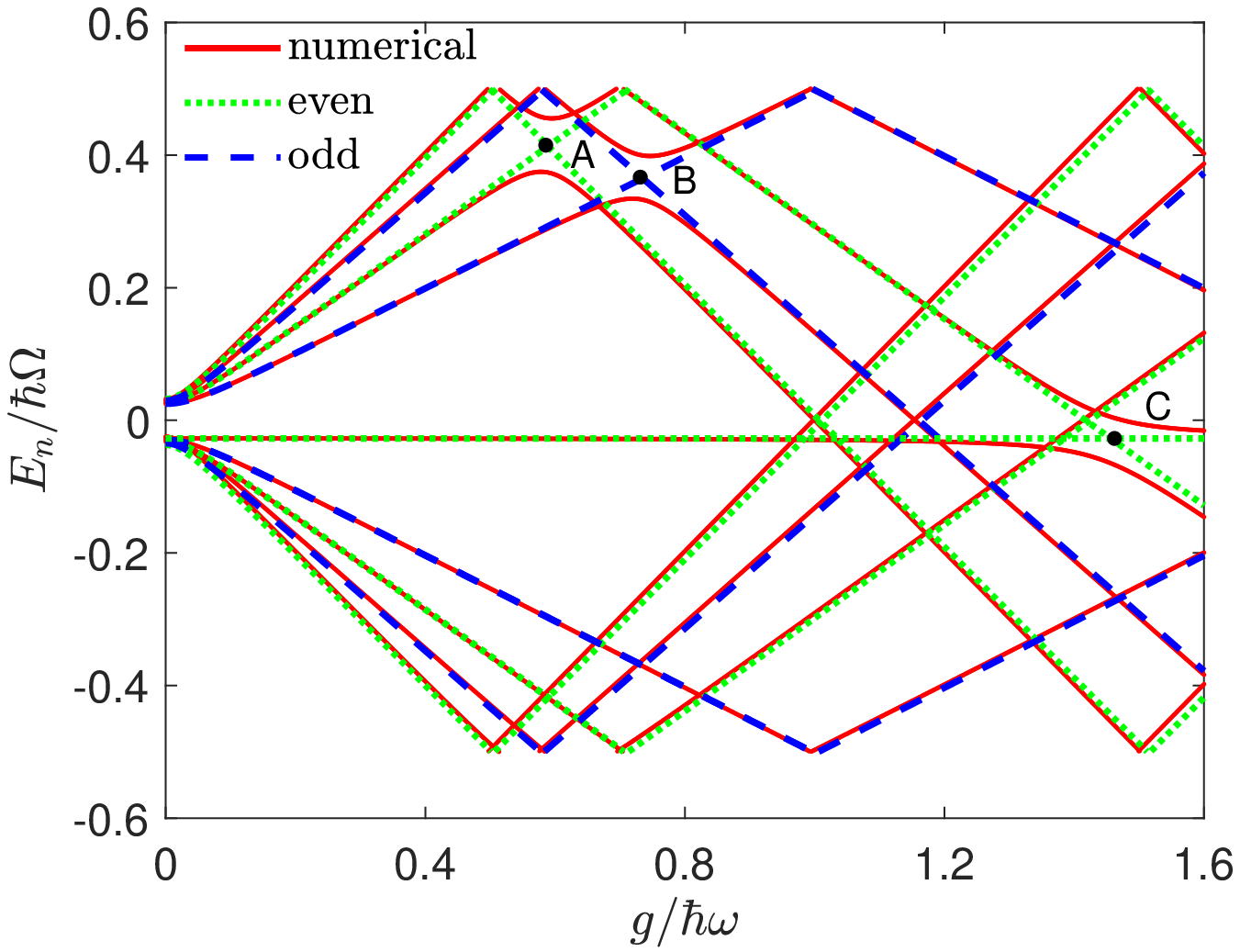}
	\caption{(Color online) The quasi-energy spectrum of the AiRM from numerical method in the extended Floquet Hilbert space (red solid) and from the effective Hamiltonian with odd parity (blue dashed) and even parity (green dotted) in the first Brillouin zone up to the photon number cutoff $n_{\rm cutoff} =3$ as a function of $g/\hbar \omega$. The set up is $\Delta = 0.1\omega$ and $g^{\prime} = 0.1 \hbar \omega$.}
	\label{Fig1}
\end{figure}

For the case that the CRTs' coupling strength $g^\prime$ is non-zero, we compare the quasi-energy spectrum result from the effective Hamiltonian with that numerically calculated in the extended Floquet Hilbert space (See Appendix B for detailed numerical scheme). The Floquet theory \cite{Eckardt2015} dictates that the solution of the Schr\"odinger equation with a time periodical Hamiltonian reads:
\begin{align}
	\Psi_n (t) = {\rm e}^{-iE_n t/\hbar} \phi_n (t), 
\end{align}
where $E_n$ is the quasi-energy and $\phi_n (t)$ is the Floquet mode. Multiplying the Floquet mode by a phase factor $\exp(i m\Omega t)$ yields the identical Floquet state but with the shifted quasi-energy $E_n +m\hbar\Omega$ with $m$ an integer. Hence the quasi-energy can be mapped into the first Brillouin zone $[-\hbar\Omega/2,\hbar\Omega/2]$ which is similar to the first Brillouin zone in the spatial lattice model, as the spectrum is invariant if translated by an integer multiple of $\hbar\Omega$. The results are shown in Fig. \ref{Fig1}, which illustrates the first few quasi-energy levels $E_{n\pm}$ with even and odd parities, i.e. the total excitation number, respectively. We fixed the CRTs coupling strength $g^{\prime} = 0.1 \hbar \omega$, and Figure 1 shows that the effective model fits the numerical result pretty well even when the rotating-wave coupling is in the deep-strong coupling regime $g\simeq\hbar\omega$. The coupling strength tends to separate the eigen-energies of the system into upper and lower branches and level crossing occurs for different parities, and avoided level crossing may occur for quasi energy with the same parity as shown below. The high-frequency expansion results, on the other hand, fails to predict the this avoid-crossing. This can be understood as follows. As we know, the conservation of total excitation number in JC model leads to the level crossing of states with the same parity, the inclusion of the counter-rotating terms in the Rabi model, either anisotropic or isotropic, however, explicitly breaks this conservation and correspondingly the level crossing of the same parity subspace cannot happen. The effective Hamiltonian employed in the Floquet calculation conserves the total excitation number $\hat{N}$ and the non-conserving terms are carried by the  kick operator $\hat{K}$. Consequently, within a given parity subspace level crossings occur in the energy spectrum as derived from the analytical results, whereas these level crossings are not present in the numerical results due to the existence of the counter-rotating terms. Explicitly the avoiding occurs at point $A$ when the positive branch for the total excitation number $N=2$ meets the negative branch for $N=4$, both of which are even parity. Similar phenomenon can be observed at point $B$ for odd parity. If we increase the photon number cutoff, more and more avoided level crossings emerge when the positive branch spectrum lines for $N$ cross the negative branch lines for $N + 2$ with the same parity. One special case is shown at point $C$, where the level for zero excitation number $E_0$ meets the even parity level for $N=2$. On the other hand, if we fix the RWTs' coupling strength $g=0.1\hbar\omega$ and vary the CWTs' $g^\prime$, the effective model coincides the numerical results well only below the regime $g^\prime\simeq 0.3\hbar\omega$ and the effective model loses its accuracy quickly for sequentially increasing $g^\prime$. In addition, we see that the detuning opens a gap $\delta E=\hbar\Delta$, defined as the difference  of upper branch and lower branch levels when $g$ approaches zero, i.e. $\delta E=\lim_{g\rightarrow0}(E_{n+}-E_{n-})$, which would be filled for large enough photon number or very strong coupling. 

\subsection{Population Dynamics}
We now turn to the study of the population dynamics of the system, in particular, we examine the dynamics with the atom initially prepared in the excited state and the field in the coherent state, which can be expressed as
\begin{align}
|\Psi_i\rangle = |\alpha\rangle \otimes |+\rangle = {\rm e}^{- \frac{|\alpha|^2}{2}}\sum_{n=0}^{\infty} \frac{\alpha^n}{\sqrt{n!}} |n\rangle\otimes|+\rangle. \label{initial}
\end{align}
with $\alpha$ is a complex parameter related to the amplitude of the coherent state. According to the Floquet theory, the time evolution of this initial state in lab frame is governed by the following time evolution operator
\begin{align}
\hat{U}(t,0) = \hat{V}(t){\rm e}^{-{\rm i}\hat{K}_{\rm eff}(t)}{\rm e}^{-\frac{\rm i}{\hbar}\hat{H}_{\rm eff} t }{\rm e}^{{\rm i}\hat{K}_{\rm eff}(0)}\hat{V}^{\dag}(0).
\end{align}
An important physical quantity is the atomic inversion, which describes the population probability difference between the $|+\rangle$ and $|-\rangle$ of the atom,
\begin{align}
W(t) = \langle\Psi(t)| \hat{\sigma}_z | \Psi(t) \rangle.
\end{align}

Both numerically and analytically result of the time evolution of the atomic inversion are shown in the Fig. 2. For clarity we plot the population dynamics for a fixed RWTs and CRTs ratio $g^\prime/g=0.5$ and a detuning $\Delta = 0.1\omega$. Due to the mean photon number in coherent state amounts to $|\alpha|^2$, for a state with $\alpha=3$ we increase our cutoff photon number to 20 to insure the calculation accuracy. The dynamics from the effective model highly coincides with the numerical result when the CRTs can not be neglected. The oscillation of the atomic inversion shows the behavior of collapse rapidly and then revival as time goes long enough. We see that with the increasing coupling strength $g$, henceforth $g\prime$, the mean atomic inversion, about which the revival signal oscillates, gradually decrease from a value in the weak coupling case, which is determined by the detuning $\Delta$, to zero, due to the involvement of RWTs and CWTs in the population probability. The collapse-revival period is shorter and shorter for increasing $g$ and the high frequency expansion describes the dynamics precisely even when the coupling strength grows above $0.2\hbar\omega$.
\begin{figure}[tbp]
	\includegraphics[width=0.5\textwidth]{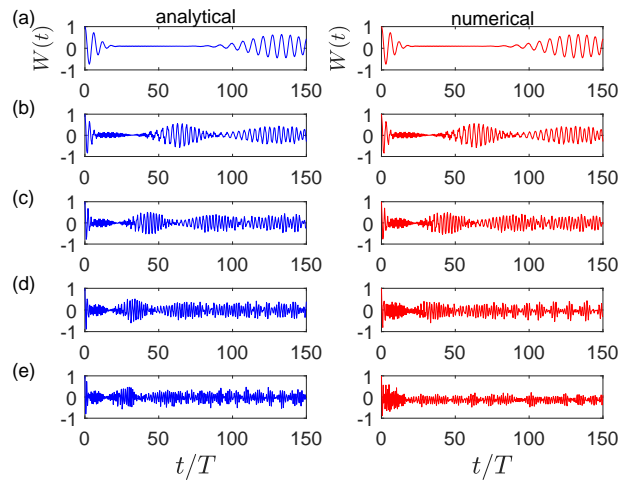}
	\caption{(Color online) The time evolution of the atomic inversion in AiRM by analytical method (blue) and numerical method (red) with detuning $\Delta = 0.1\omega$, the amplitude of the coherent sate $\alpha=3$, the fixed ratio between RWTs and CRTs $g^\prime/g=0.5$, and the cutoff photon number $n_{\rm cutoff} = 20$. The time scale is $T = 2\pi/\Omega$.  (a)$g=0.05\hbar\omega$, (b)$g=0.1\hbar\omega$, (c)$g=0.15\hbar\omega$, (d)$g=0.2\hbar\omega$ , (e)$g=0.25\hbar\omega$.}
	\label{Fig2}
\end{figure}

\section{Asymmetric Rabi model}

\subsection{Formalism and high-frequency expansion}

Another model we choose to illustrate the power of high-frequency expansion is the asymmetric quantum Rabi model (AsRM) \cite{Larson2013,LiuMX2017} with Hamiltonian written as
\begin{align}
\hat{H}_{\rm{AsRM}} = \frac{1}{2}\hbar\omega_{0}\hat{\sigma}_{z}+\varepsilon\hat{\sigma}_x+\hbar\omega\hat{a}^{\dag}\hat{a}+g(\hat{a}^{\dag}+\hat{a})(\hat{\sigma}_++\hat{\sigma}_-),
\end{align}
where a bias external field $\varepsilon$ is applied along the $x-$axis, which sometimes is regarded as the intrinsic transition strength of the atom \cite{Wakayama2017}, whereas other parameters are kept the same as in the anisotropic Rabi model in previous section. Here the atom-field coupling is chosen as the Rabi type with equal coupling strength of RWTs and CRTs. This $\varepsilon$ term also breaks the $Z_2$ symmetry of the quantum Rabi model, however, provides more realistic description of the circuit QED experiments employing flux qubits than the Rabi model itself \cite{Niemczyk2010}. After similar steps we applied earlier, the Hamiltonian in the rotating frame reads
	\begin{align}
		\hat{H}_{\rm{AsRM}}^{\rm rot}(t)&=\frac{\hbar\Delta}{2}\hat{\sigma}_z +g(\hat{a}^{\dag}\hat{\sigma}_{-}+ \hat{a}\hat{\sigma}_{+})\notag\\
		&+\varepsilon({\rm e}^{{\rm i}\omega t}\hat{\sigma}_{+}+{\rm e}^{-{\rm i}\omega t}\hat{\sigma}_{-})\notag\\
		&+g({\rm e}^{{\rm i}2\omega
			t}\hat{a}^{\dag}\hat{\sigma}_{+}+{\rm e}^{-{\rm i}2\omega t}\hat{a}\hat{\sigma}_{-}).
		\label{HAsRM}
	\end{align}

Clearly there exist two driving terms with frequencies $\omega$ and $2\omega$ and we choose the driving frequency as $\Omega = \omega$. The first few terms of the effective Hamiltonian $\hat{H}_{\rm eff}$ are calculated as
\begin{align}
H_{\rm eff}^{(0)} &= \frac{1}{2}\hbar\Delta\hat{\sigma}_z+g(\hat{a}^{\dag}\hat{\sigma}_{-}+\hat{a}\hat{\sigma}_{+}),\label{Heff0}\\
H_{\rm eff}^{(1)} &= \frac{1}{\hbar\Omega}[\varepsilon^{2}\hat{\sigma}_{z}+\frac{g^2}{2}(\hat{a}^{\dag}\hat{a}\hat{\sigma}_z-\hat{\sigma}_-\hat{\sigma}_+)],\label{Heff1}\\
H_{\rm eff}^{(2)} &=-\frac{1}{(\hbar\Omega)^{2}}\{\hbar\Delta\varepsilon^2\hat{\sigma}_z+2g\varepsilon^2(\hat{a}^{\dag}\hat{\sigma}_-+h.c.)\notag\\
&+\frac{1}{4}[\hbar\Delta g^2(\hat{a}^{\dag}\hat{a}\hat{\sigma}_z-\hat{\sigma}_-\hat{\sigma}_+)+g^3(\hat{a}^{\dag }\hat{a}\hat{a}^{\dag }\hat{\sigma}_-+h.c.)]\} \label{Heff2}
\end{align}
with the corresponding kick operator $\hat{K}_{\rm eff}(t)$
\begin{align}
K_{\rm eff}^{(1)}(t) &= \frac{1}{2{\rm i}\hbar \Omega}\left(2\varepsilon\hat{\sigma}_+{\rm e}^{{\rm i}\Omega t}+g\hat{a}^{\dag}\hat{\sigma}_+{\rm e}^{{\rm i}2\Omega t}-h.c.\right),\\
K_{\rm eff}^{(2)}(t) &=\frac{1}{4{\rm i}(\hbar\Omega)^{2}}[\varepsilon(7 g\hat{a}^{\dag}\hat{\sigma}_z-4\hbar\Delta{\sigma}_+){\rm e}^{{\rm i}\Omega t}\notag\\
&+g(ga^{\dag2}{\sigma}_z-\hbar\Delta \hat{a}^{\dag}{\sigma}_+){\rm e}^{{\rm i}2\Omega t}-h.c.].
\end{align}
As we can see, the zeroth order effective Hamiltonian (\ref{Heff0}) takes the same JC form as in the anisotropic Rabi model and again plays a major role in the effective Hamiltonian. The bias $\varepsilon$ will bring direct bearing on the effective Hamiltonian by means of the quasi-level difference as it induces an additional $\sigma_z$ term, and together with the coupling parameter $g$ the driving terms contribute to the atom-field interaction up to $(l+1)$-photon process for the $l$-th order correction. The kick operators responsible for the dynamics are classified into two terms with a two-frequency driving with frequencies $\Omega$ and $2\Omega$, controlled jointly by the bias $\varepsilon$ and coupling parameter $g$.

\subsection{Quasi-energy and eigenstates}

The diagonalization of the effective Hamiltonian in the basis $|1\rangle = |n,+\rangle$ and $\ |2\rangle = |n+1,-\rangle$ gives the quasi-energy spectrum. The sole state with zero excitation number $| \Phi_0 \rangle=|0, - \rangle$ is the same as in AiRM model with quasi energy
\begin{align}
E_0 & = -\frac{\hbar\Delta}{2}-\frac{\varepsilon^2+g^2/2}{\hbar\Omega} + \frac{\hbar\Delta(\varepsilon^2+g^2/4)}{(\hbar\Omega)^2},
\end{align}
while the expansion of the quasi energy for those with non-zero total excitation number are
\begin{align}
		E_{n\pm}^{(0)} &= \pm\frac{\Omega_{R}}{2},\\
		E_{n\pm}^{(1)} &=\frac{1}{2\hbar\Omega}\left\{-g^2\pm\frac{\hbar\Delta \Omega_\varepsilon^2}{\Omega_{R}}\right\},
\end{align}
and
\begin{widetext}
	\begin{align}
		E_{n\pm}^{(2)}  &= \frac{1}{(2\hbar\Omega)^2}\left\{\hbar\Delta g^{2}\pm\frac{\Omega_\varepsilon^4-8\Omega_\varepsilon^2g^2(n+1)+6g^4(n+1)^2-(\hbar\Delta)^2 [2\Omega_\varepsilon^2+\Omega_\varepsilon^2/\Omega_R^{ 2}- g^2(n+1)]}{\Omega_{R}}\right\},
	\end{align}
\end{widetext}
where we have defined a bias-related frequency $\Omega_\varepsilon = \sqrt{2\varepsilon^2+g^2(n+1)}$ in a similar way as the Rabi frequency $\Omega_R$. And the corresponding eigenvectors take the same form as in anisotropic Rabi model in eq. (\ref{phinpm}), with the coefficients given by
\begin{align}
C_{n\pm}^{(0)} &= \frac{\hbar\Delta\pm\Omega_R}{2g\sqrt{n+1}},\\
C_{n\pm}^{(1)} &=\frac{1}{\hbar\Omega}\frac{1}{g\sqrt{n+1}}\left(\varepsilon^2+\frac{g^2(n+1)}{2}\pm\frac{\hbar\Delta\Omega_\varepsilon^2}{2\Omega_R}\right),\\
C_{n\pm}^{(2)} &= \frac{g\sqrt{n+1}}{(\hbar\Omega)^2}\left\{-\frac{\hbar\Delta}{8}\mp\frac{(\hbar\Delta)^2}{8\Omega_R} \pm\frac{\Omega_\varepsilon^4}{\Omega_R^{3}}\right\}.
\end{align}
\begin{figure}[tbp]
	\includegraphics[width=0.5\textwidth]{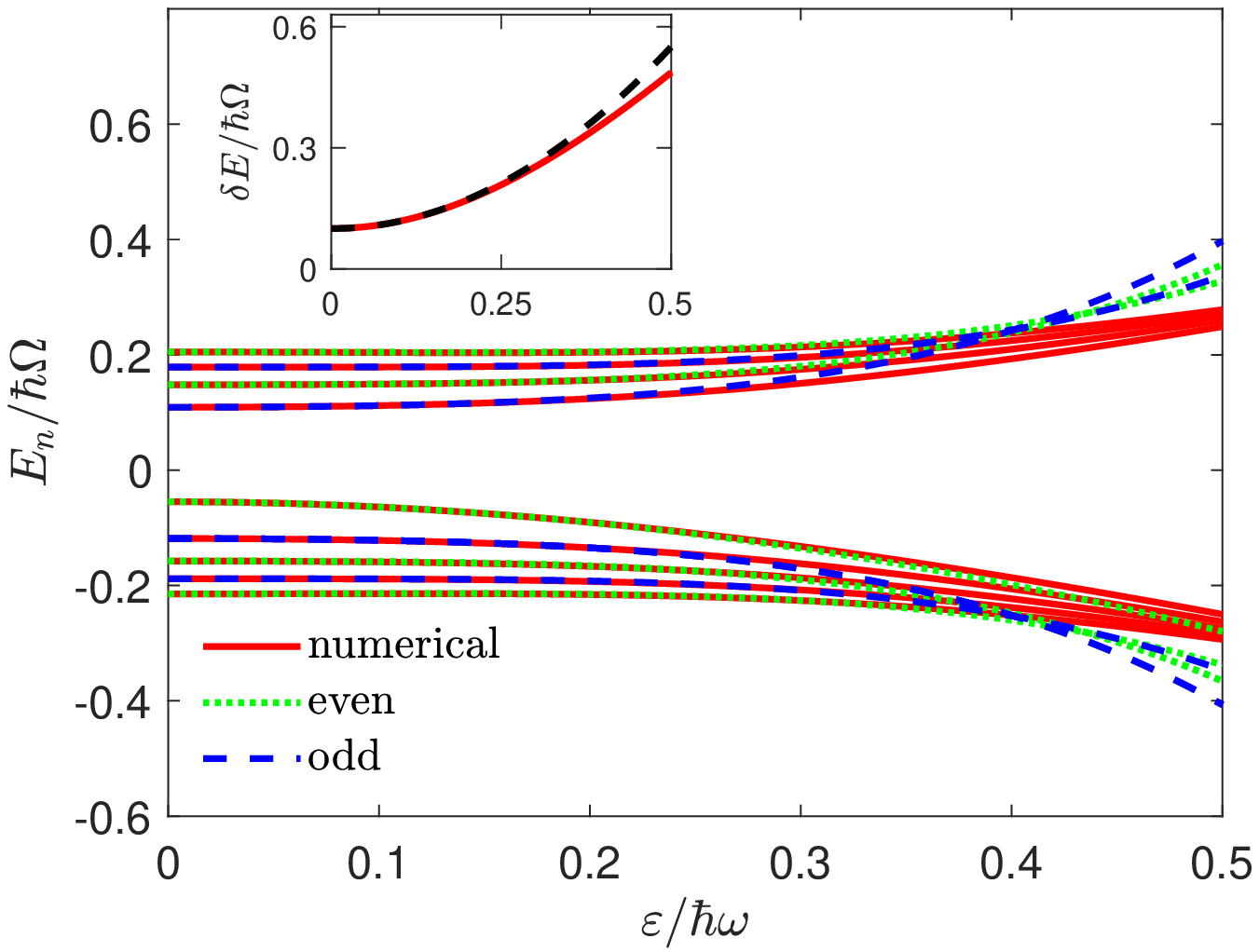}
	\caption{(Color online) The quasi-energy spectrum for $n=0,1,2,3$ of the AsRM from numerical method in the extended Floquet Hilbert space (red solid) and from the effective Hamiltonian with odd parity (blue dashed) and even parity (green dotted) in the first Brillouin zone calculated up to the photon number cutoff $n_{\rm cutoff} =10$ as a function of $g/\hbar \omega$. The set up is $\Delta = 0.1\omega$ and the coupling strength $g = 0.1\hbar \omega$. Inset: The numerical (red solid) and analytical (black dashed) results of the gap $\delta E$ dependence on the bias field $\varepsilon$ at the limit $g \rightarrow 0$.}
	\label{Fig3}
\end{figure}

Numerical result of the quasi energy spectrum and that from the effective model are shown in Fig. \ref{Fig3}. Here the photon number cutoff is taken as $n_{\rm cutoff}=10$ in order to assure the accuracy of the levels for $n=0,1,2,3$ for a comparison of the matrix diagonalization result in the extended Floquet Hilbert space and the high frequency expansion result for even and odd parities, respectively. We again present the spectrum in the first Brillouin zone, $[-\hbar\Omega/2,\hbar\Omega/2]$. To see more clearly the role played by the bias field $\varepsilon$, both the detuning and the coupling strength are fixed to a moderate value of $0.1\hbar \omega$, and similar with the case of the AiRM, the effective model provides an efficient tool for the treatment of the quasi energy spectrum in a rather wide parameter regime of the bias field up to $0.3\hbar \omega$. We see the bias field tends to cluster the upper and lower branches $E_{n\pm}$ into two bundles, although the concentration point given by the effective model is earlier than the numerical results. Avoided level crossing never happens here due to the asymmetric structure of the AsRM model, as the bias breaks the parity symmetry in the standard Rabi model.

As in the AiRM, the detuning opens a gap $\delta E$ in the quasi energy spectrum. What different here is that in the case of AsRM this gap is bias dependent. For concreteness, in the inset of Fig. \ref{Fig3} we plot this gap as a function of the bias field $\varepsilon$ in the limit $g \rightarrow 0$. Evidently the effective model already capture the main feature of this gap. From the first few terms of the effective Hamiltonian eqs. (\ref{Heff0}) to (\ref{Heff2}) one can easily see that the diagonal terms in the form of $\sigma_z$ will determine the gap dependence on the bias as 
\begin{align}
	\delta E=\hbar\Delta+2(\hbar\Omega-\hbar\Delta)\left(\frac{\varepsilon}{\hbar\Omega}\right)^2.
\end{align}
This quadratic denpendence fits the numerical result for a bias field $\varepsilon$ up to $0.4 \hbar \omega$. Note here the high frequency expansion request a more strict condition for the atom-field coupling $g \simeq 0.1\hbar\omega$, to make sure that the driven frequency $\Omega$ do dominate the energy scale because the driven frequency here is half of that in AiRM. But for the low-energy, such as the photon number $n$ is zero, the effective model fits well even in the deep-strong coupling regime as we mentioned before.


\begin{figure}[tbp]
	\includegraphics[width=0.5\textwidth]{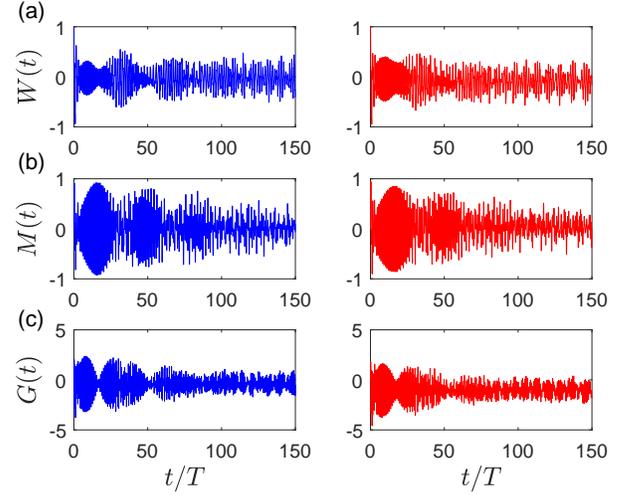}
	\caption{(Color online) The long time evolution of some physical observables by analytical method (blue) and numerical method (red) in the same panel with the set up that coupling strength $g=0.1\hbar\omega$, detunning $\Delta=0.1\omega$, and the bias strength $\varepsilon=0.1\hbar\omega$. Shown in the panels are the expectation value of (a) the atomic inversion, (b) the transverse magnetization, and (c) the atom-field correlation. }
	\label{Fig4}
\end{figure}


\subsection{Driving Dynamics and Fourier Spectrum}

\begin{figure}[tbp]
	\includegraphics[width=0.5\textwidth]{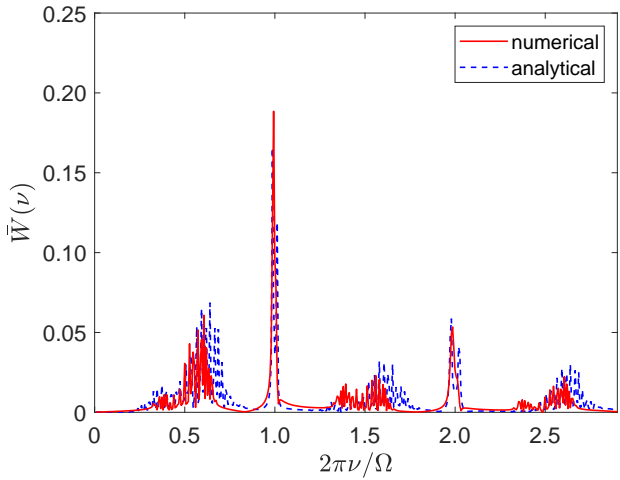}
	\caption{(Color online) The Fourier frequency spectrum analysis of the time-evolution of the atomic inversion, where we choose the parameters $g=0.1\hbar\omega$ $\varepsilon = 0.3\hbar\omega$ $\Delta=0.1\omega$.}
	\label{Fig5}
\end{figure}

The physical observables, such as the atomic inversion $W(t)$ introduced in last section, evolve with time and we are interested in the driving dynamics and the  steady oscillation properties for long enough driving time. For the AsRM, it is of interest to consider the magnetization $M(t)$ induced by the transverse bias field $\varepsilon$, and the atom-field correlation $G(t)$ mediated by the coupling parameter $g$. The initial state (\ref{initial}) is chosen the same as that in the previous model and the definition of the latter two physical observables are given by
\begin{align}
	M(t) &= \langle\Psi(t)| \hat{\sigma}_x | \Psi(t) \rangle,\\
	G(t) &= \langle\Psi(t)| (\hat{a}^\dagger+\hat{a})\hat{\sigma}_x | \Psi(t) \rangle.
\end{align}
As we can see in Fig. \ref{Fig4}, for a short time the evolution of all these observables exhibits a collapse and revival phenomenon which takes the form of wave packets. The amplitude of the wave packet decreases and the oscillation tends to be stable as time goes by so that the wave packet can not be observed any more. With suitable choice of the system parameters, the analytical method is sufficiently accurate to describe the system evolution by comparing with the numerical results. To further understand the nature of these oscillation, we apply the Fourier spectrum analysis to  extract the frequency in the oscillation of these three observables. The Fourier transform is the fundamental technique of Fourier analysis, and it decomposes the original data into its frequency components, which is often referred to as the frequency spectrum. The Fourier transform is represented as
\begin{align}
	\bar{F}(\nu)= \int_{0}^{+\infty} {\rm d}t F(t) {\rm e}^{-{\rm i}2\pi\nu t},
\end{align}
where $\bar{F}(\nu)$ is the output spectrum that is a function of frequency $\nu$, $F(t)$ is the input data that is a function of time $t$. The driving dynamics of three observables are similar and we take the atomic inversion $W(t)$ as an example. The frequency spectrum of atomic inversion shows the feature of two-frequency driving behavior in Fig. \ref{Fig5}, i.e. both the analytical and numerical results indicate that the fundamental frequency is located at $\Omega$ and the second harmonics is located at $2\Omega$ as expected. A relatively large external bias field $\varepsilon=0.3\hbar \omega$ is to enhance the peak value at the fundamental frequency as the bias dominate the oscillation $e^{\pm {\rm i} \omega t}$ in the rotating frame Hamiltonian (\ref{HAsRM}).  The involvement of many-photon Fock states in the coherent state leads to the broadening of the spectral functions at both the fundamental frequency and the second harmonics, as well as the complicated oscillation around the inevitable frequency mixing at $0.5, 1.5$ and $2.5 \Omega$. The double and triple revival sequences for the two- and three-qubit systems have been found in the probability of finding all qubits in the initial $|+\rangle$ state as a consequence of having two or three Rabi frequencies \cite{Agarwal2012,Mao2016}. However, the second harmonics here originates from a rather different mechanics as the effective Hamiltonian for AsRM is basically a two-frequency driving system.

\begin{figure}[tbp]
	\includegraphics[width=0.5\textwidth]{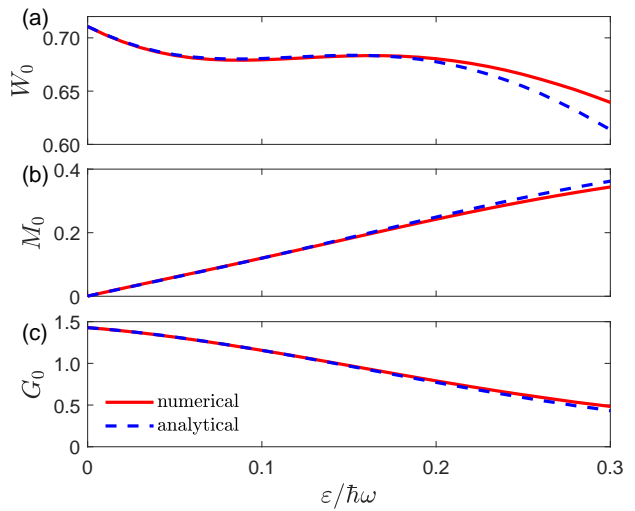}
	\caption{(Color online) The time-averaged value of atomic inversion $W_0$, transverse magnetization $M_0$, and atom-field correlation $G_0$ over the 150 driving periods as a function of bias field $\varepsilon$ for $g=0.01\hbar\omega$ and $\Delta=0.1\omega$.}
	\label{Fig6}
\end{figure}

The other observation is that for long enough time the system is driven into a steady state. It is of interest how the bias field could control the system and what time-averaged value of the observables would be reached. We define the time-averaged values of atomic inversion $W_0$, transverse magnetization $M_0$, and atom-field correlation $G_0$ as the average over the 150 driving periods, and show their dependence on the varying $\varepsilon$. Fig. \ref{Fig6} shows the numerical and high frequency expansion results for a small coupling parameter $g$, to focus on the controllability of the bias field. The time-averaged value of atomic inversion $W_0$ experiences a competition between detuning $\Delta$ and the coupling strength $g$. The effect of detuning is equivalent to hindering the atomic inversion, while the effect of $g$ is to induce transition between upper and lower energy levels. When $g$ and $\varepsilon$ are both small, they fight against detuning together, making the population begin to reverse. When $\varepsilon$ increases to a certain extent, it will start to compete with $g$. Thus we see a regime where $W_0$ is nonetheless slightly increased. The atom will start to reverse again until the effect of $g$ is completely eliminated and $\varepsilon$ takes the dominant role. On the other hand, the time-averaged value of the transverse magnetization $M_0$ increases with applied bias field linearly as expected, whereas the bias field serves to destroy the correlation between atom and field. In other parameter regime the competition between three parameters $g$, $\varepsilon$ and $\Delta$ remains, even leading to negative atom-field correlation, which is not shown in Fig. \ref{Fig6}. The effective model proves to be very accurate till $\varepsilon \sim 0.2 \hbar \omega$ thus provides a powerful tool in estimating the dynamics and the time-averaged values of these observables.

Finally we discuss regime of validity of our high-frequency expansion scheme used in this paper. First of all, the high frequency expansion need the driving frequency $\Omega$ to be large for both AiRM and AsRM. Secondly, for a fixed RWT coupling $g=0.1\hbar\omega$ the effective AiRM model works reasonably well in a range of detuning $-1 < \Delta/\omega < 2$, i.e. either blue $\Delta<0$ or red detuning $\Delta>0$, provided that the CRT or the bias is below $0.3\hbar\omega$. For fixed CRTs coupling strength $g^\prime=0.1\hbar\omega$ the effective AiRM model fits the numerical result surprisingly well even for the RWT coupling $g\sim1.5\hbar\omega$ in the whole range of detuning. This is due to that fact that the $g^\prime$ term is a driving term while $g$ term is not - the relatively larger $g$ only gives a boost to the zeroth-order effective Hamiltonian in (\ref{H0}). The valid regime for $g^\prime=0.1\hbar\omega$ is thus a rectangle in the $\Delta-g$ plane, i.e $-1<\Delta/\omega <2$, $0<g<1.5\hbar\omega$, as shown in the Figure 8 in reference \cite{Hausinger2010}. If we increase the driving term $g^\prime$, the left up corner will first become invalid and for a strong enough driving term $g^\prime=0.25\hbar\omega$ the method is inaccurate also in a small area near the resonance even for small $g$. Further increase of $g^\prime$ will totally invalidate the high frequency expansion. From the view point of dynamics, it is clearly that close to resonance the analytical results match the numerics for fixed $g^\prime/g=0.5$ provided that $g^\prime$ is below $0.125\hbar\omega$. For fixed $g=\hbar\omega$, we also find good match for $g^\prime$ up to $0.125\hbar\omega$. For AsRM, we fixed the couplings in the ultra-strong regime, i.e. $g=0.1\hbar\omega$, and a small detuning $\Delta=0.1\omega$, and find the high frequency expansion is valid for a relative large bias $\varepsilon =0.3\hbar\omega$. 

It is also necessary to compare our results with other analytical methods presented in the literature. Generalized rotating wave approximation (GRWA) \cite{Irish2007} and the Van Vleck perturbation (VVP) theory \cite{Hausinger2008,Hausinger2010} are both valid in the case of large blue detuning, $-1<\Delta/\omega<-0.4$, from the weak to the deep-strong coupling $0<g<1.5\hbar\omega$. The GRWA is preferable to VVP at weak coupling, in particular close to resonance and red detuning. On the contrary, VVP works better at strong coupling strengths. High frequency expansion perfectly fills in the blank left by these two methods on the right part of the red detuning regime, for a ultra-strong CRT coupling parameter up to $g^\prime=0.25\hbar\omega$.

\section{Conclusion}
In conclusion, we transformed two extended Rabi models, i.e. the AiRM and AsRM, into the rotating frame, and regard them as the periodically driven models. By applying the Floquet theory and the high-frequency expansion, we obtained the effective model Hamiltonian and the quasi-energy spectrum both analytically and numerically. For AiRM, the effective model agrees pretty well with the numerical diagonalization in the extended Floquet Hilbert space for $g$ up to  $2 \hbar\omega$ for a CRT coupling in the ultrastrong coupling regime $g^\prime=0.1 \hbar \omega$. The effective model fails to predict the avoided level crossing occurred in the same parity due to its conservation of total excitation number. The population dynamics governed by the effective model, however, is accurate enough for a fixed ratio of the RWTs and CRT coupling strength $g^\prime/g=0.5$. For AsRM, the quasi energy spectrum is found to be clustered into two bundles by the bias field, which breaks the parity symmetry in the Rabi model. In both cases, the detuning opens a gap in the quasi energy spectrum illustrated in the first temporal Brillouin zone, which is exactly the detuning energy in the AiRM and depends quadratically on the bias field in the AsRM. The driving dynamics of several observables are studied by means of the Fourier analysis and the two-frequency driving nature is manifested in the frequency spectrum. The time-averaged value of these oscillation may be controlled by the bias field, while a competition with detuning and atom-field coupling is expected to provide more versatile means to manipulate the driving dynamics. The Floquet method in the extended Rabi models provides an alternative tool in the study of interaction between atom and light and is readily applied to more sophisticated models where more qubits, more cavity modes, or many-body interaction are involved.

\addcontentsline{toc}{chapter}{Appendix A: Derivation of the effective Hamiltonian}
\section*{Appendix A: Derivation of the effective Hamiltonian}
The formula for high-frequency expansion of the effective Hamiltonian up to the second order can be written as \cite{Bukov2015, Goldman2014}
\begin{align}
\hat{H}_{\rm eff}^{(0)} &= H_{0},\\
\hat{H}_{\rm eff}^{(1)} &=\frac{1}{\hbar\Omega}\sum_{l=1}^{\infty}\frac{[H_{l},H_{-l}]}{l},\\
\hat{H}_{\rm eff}^{(2)} &= \frac{1}{(\hbar\Omega)^{2}}\sum_{l\neq0}\left(\frac{[[H_{l},H_{0}],H_{-l}]}{2l^{2}}\right.\notag\\
&\left.+\sum_{l^{\prime}\neq0,-l}\frac{[[H_{l},H_{l^{\prime}}],H_{-(l+l^{\prime})}]}{3l(l+l^{\prime})}\right).
\end{align}
For $\hat{K}_{\rm eff}^{(n)}(t)$ up to order $\frac{1}{\Omega^{2}}$, we can get
\begin{align}
\hat{K}_{\rm eff}^{(1)}(t) &= \frac{1}{{\rm i}\hbar\Omega}\sum_{l\neq0}\frac{H_{l}{\rm e}^{{\rm i}l\Omega t}}{l},\\
\hat{K}_{\rm eff}^{(2)}(t) &=\frac{1}{{\rm i}(\hbar\Omega)^{2}}\sum_{l\neq 0}\frac{[H_{l},H_{0}]{\rm e}^{{\rm i}l\Omega t}}{l^{2}}\notag\\
&+\frac{1}{2{\rm i}(\hbar\Omega)^{2}}\sum_{l\neq0}\sum_{l^{\prime}\neq0,-l}\frac{[H_{l},H_{l^{\prime}}]{\rm e}^{{\rm i}(l+l^{\prime})\Omega t}}{l(l+l^{\prime})},
\end{align}
where $H_l$ is the Fourier expansion coefficients of $l$-th order,
\begin{align}
H_l = \frac{1}{T}\int_{0}^{T}\hat{H}(t){\rm e}^{-l\Omega t} {\rm d}t.
\end{align}
For the anisotropic Rabi model \eqref{Hrot}, we have
\begin{align}
H_0 &= \frac{1}{2}\hbar\Delta\hat{\sigma}_{z}+g(\hat{a}^{\dag}\hat{\sigma}_{-}+\hat{a}\hat{\sigma}_{+}),\\
H_{-1} &= g^{\prime}\hat{a}\hat{\sigma}_{-},\\
H_{1} &= g^{\prime}\hat{a}^{\dag}\hat{\sigma}_{+},
\end{align}
while for the asymmetric Rabi model \eqref{HAsRM}, we can get
\begin{align}
H_0 &= \frac{1}{2}\hbar\Delta\hat{\sigma}_{z}+g(\hat{a}^{\dag}\hat{\sigma}_{-}+\hat{a}\hat{\sigma}_{+}),\\
H_{-1} &= \varepsilon\sigma^{-},\qquad H_{1} = \varepsilon\sigma^{+},\\
H_{-2} &= g\hat{a}\sigma^{-},\qquad H_{2} = g\hat{a}^{\dag}\sigma^{+}.
\end{align}
By plugging them into the formula above, we can get the effective Hamiltonian and effective kick operators for each model.
\addcontentsline{toc}{chapter}{Appendix B: Numerical method for quasi-energy spectrum}
\section*{Appendix B: Numerical method for quasi-energy spectrum}

The numerical procedure is based on the block diagonalization of the quasi energy operator in the extended Floquet Hilbert space by means of degenerate perturbation theory in Ref. \cite{Eckardt2015}. Let us consider an extended Floquet Hilbert space $\mathscr{F}$, which is given by the direct product of the state space $\hat{H}$ and the space of square-integrable $T$-periodically time-dependent function $\mathcal{L}_T$. In this case, a complete set of orthonormal basis states $|\alpha m(t)\rangle\rangle$ of space $\mathscr{F}$ can be constructed by combining a complete set of orthonormal basis states of $\hat{H}$,  $| \alpha \rangle = | n, \pm \rangle$, with the complete set of time-periodic functions ${\rm e}^{{\rm i}m\Omega t}$ labeled by the integer $m$. In matrix form, we can get
\begin{align}
|\alpha m(t)\rangle\rangle =| n, \pm \rangle  {\rm e}^{{\rm i}m\Omega t}.
\end{align}
Using the definition of the scalar product in the extended Floquet Hilbert space, we can get the matrix elements $\mathcal{F}_{m^\prime m}^{\alpha^\prime\alpha}=\langle\langle\alpha^{\prime} m^{\prime}|\hat{\mathcal{F}}|\alpha m\rangle\rangle$ of the Floquet operator $\hat{\mathcal{F}}=\hat{H}^{\rm rot}(t)-{\rm i}\hbar \frac{\partial }{\partial t}$ with respect to the basis $|\alpha m\rangle\rangle$,
\begin{align}
\mathcal{F}_{m^\prime m}^{\alpha^\prime\alpha} &= \frac{1}{T}\int_0^{T}{\rm d}t\  {\rm e}^{-{\rm i}m^\prime\Omega t}\langle\alpha^{\prime}|\hat{H}^{\rm rot}(t) -{\rm i}\hbar \frac{\partial}{\partial t}|\alpha\rangle{\rm e}^{{\rm i}m\Omega t}\notag\\
&=\langle\alpha^\prime|H_{m^\prime-m}|\alpha\rangle +\delta_{m^\prime m}\delta_{\alpha^\prime\alpha}m\hbar\Omega,
\end{align}
where
\begin{align}
H_{m^\prime-m} = \frac{1}{T}\int_0^{T}{\rm d}t{\rm e}^{-{\rm i}(m^\prime-m)\Omega t}\hat{H}^{\rm rot}(t).
\end{align}
Quasi-energy is the eigenvalue of this infinite matrix.
\subsection*{B1. Anisotropic Rabi model}
For the anisotropic Rabi model, the Floquet operator matrix can be expressed as
\begin{align}
\begin{pmatrix}
\ddots&\vdots&\vdots&\vdots&\vdots&\ddots\\
\cdots&H_0-\hbar\Omega&H_{-1}&0&0&\cdots\\
\cdots&H_1&H_0&H_{-1}&0&\cdots\\
\cdots&0&H_1&H_{0}+\hbar\Omega&H_{-1}&\cdots\\
\cdots&0&0&H_{1}&H_{0}+2\hbar\Omega&\cdots\\
\ddots&\vdots&\vdots&\vdots&\vdots&\ddots
\end{pmatrix},
\end{align}
where
\begin{align}
H_{m^\prime-m} = \begin{cases}
g^{\prime}\hat{a}\hat{\sigma}_{-},&m^\prime-m=-1\\
\frac{1}{2}\hbar\Delta\hat{\sigma}_{z}+g(\hat{a}^{\dag}\hat{\sigma}_{-}+\hat{a}\hat{\sigma}_{+}),&m^\prime-m=0\\
g^{\prime}\hat{a}^{\dag}\hat{\sigma}_{+},&m^\prime-m=1.
\end{cases}
\end{align}
This infinite matrix cannot be solved analytically, so we have to cut off the $m$-index and the photon number $n$ and diagonalize the matrix numerically. The maximum of these two numbers are $m_{\rm max}=10$ and $n_{\rm max} = 4$.
\subsection*{B2. Asymmetric Rabi model}
For asymmetric Rabi model, the Floquet operator matrix can be expressed as
\begin{align}
\begin{pmatrix}
\ddots&\vdots&\vdots&\vdots&\vdots&\ddots\\
\cdots&H_0-\hbar\Omega&H_{-1}&H_{-2}&0&\cdots\\
\cdots&H_1&H_0&H_{-1}&H_{-2}&\cdots\\
\cdots&H_{2}&H_1&H_{0}+\hbar\Omega&H_{-1}&\cdots\\
\cdots&0&H_{2}&H_{1}&H_{0}+2\hbar\Omega&\cdots\\
\ddots&\vdots&\vdots&\vdots&\vdots&\ddots
\end{pmatrix},
\end{align}
where
\begin{align}
H_{m^\prime-m} = \begin{cases}
g\hat{a}\hat{\sigma}_{-},&m^\prime-m=-2\\
\varepsilon\hat{\sigma}_-,&m^\prime-m=-1\\
\frac{1}{2}\hbar\Delta\hat{\sigma}_{z}+g(\hat{a}^{\dag}\hat{\sigma}_{-}+\hat{a}\hat{\sigma}_{+}),&m^\prime-m=0\\
\varepsilon\hat{\sigma}_+,&m^\prime-m=1\\
g\hat{a}^{\dag}\hat{\sigma}_{+},&m^\prime-m=2.
\end{cases}
\end{align}
The cut-off condition we choose is the same as in the anisotropic Rabi model.
\addcontentsline{toc}{chapter}{Acknowledgment}
\section*{Acknowledgment}
The authors are grateful to Dr. C.-M. Dai for illuminating discussions on Magnus expansion. This work is supported by the National Natural Science Foundation of China (Grant No. 12074340) and the Science Foundation of Zhejiang Sci-Tech University (ZSTU) under Grant no. 20062098-Y.

\end{document}